\documentstyle[aps,prl]{revtex}

\begin{document}
\draft
\title{Quantum theory of excess noise}
\author{P.\ J.\ Bardroff and S.\ Stenholm}
\address{Department of Physics, Royal Institute of
  Technology (Kungl.\ Tekniska H\"ogskolan),
  Lindstedtsv\"agen 24, S-10044 Stockholm, Sweden
}
\date{\today}
\maketitle
\begin{abstract}
We analyze the excess noise in the framework of the conventional
quantum theory of laser-like systems. Our calculation is conceptually
simple and our result also shows a correction to the semi-classical
result derived earlier.
\end{abstract}
\pacs{PACS numbers:  42.50.Lc, 42.55.Ah}

In attenuators, amplifiers and laser systems, the noise of the signal
is increased by the interaction with a reservoir. In usual laser
systems this is reflected in the formula for the Schawlow-Townes
linewidth \cite{schawlow,scully,lax} which provides a simple relation
between the linewidth and the gain or loss of the system. Under
certain conditions, however, the noise entering from the reservoir can
exceed this minimum amount by a large factor --- the so-called
excess-noise or Petermann factor. The appearance of this factor was
predicted by Petermann \cite{peter} in the context of gain-guided
semiconductor lasers. Within a semi-classical theory \cite{semi}, this
concept was later generalized to other systems in quantum electronics.

However, for a general system a complete quantum
mechanical derivation is still lacking.
Most of the considerations so far include the quantum noise
properties ad hoc into an otherwise classical theory.
Only a few simple systems have been discussed quantum mechanically
in a rigorous way \cite{gold,squant}.

During recent years, many interesting experiments have been carried
out: The first experiments used cavities with large output coupling
\cite{output} enhancing the noise by a few times only. However,
geometrically unstable laser cavities show excess-noise factors up to
a few hundreds for both solid state lasers \cite{semicond} and gas
lasers \cite{gas}. Also the coupling of the polarizations of the laser
light \cite{polar} and the insertion of a small aperture
\cite{aperture} can lead to large excess noise.

We derive the noise properties of an amplifier starting from the usual
quantum mechanical description of the radiation field. First we derive
a multi-mode master equation for the amplification. This is done in a
way analogous to the maser theory assuming a reservoir of excited
atoms \cite{scully}. Then we are able to define quasi modes of the
total system which show noise enhanced by the so-called $K$-factor.
Our calculation is conceptually simple and transparent; it applies to
all linear systems. We derive a correction to the semi-classical
theory of Ref.~\cite{semi}. We subsequently generalize our result by
including damping which allows us to apply it to the case of a laser
below or up to threshold. We explain the main features of the excess
noise and discuss its properties.
For the experimentally relevant cases, we show that 
our calculations justify the use of the ordinary
semi-classical treatment of excess noise. 
Thus we refrain from providing any explicit physical realization of
our abstract model. 
Such a calculation would, in all essential features, reproduce the
calculations carried out in the various model systems discussed in the
literature \cite{gold}. 
In this paper we do not include the
effects of nonlinear saturation which needs to be considered
separately.

Here we briefly recall the standard expressions of the quantized
electromagnetic field in order to define our notation.
We use orthonormal real mode functions $u_n(x)$ of the electromagnetic
field with frequency $\omega_n$ which fulfill the boundary conditions
for the given configuration in the whole ``universe'' and satisfy the
orthonormality relation
\begin{equation}
  \frac{1}{V}\int d^3x\,u_n(x)u_m(x)=\delta_{nm},
  \label{eq:orthu}
\end{equation}
where $V$ is the volume of the whole space.
Note that the mode function $u_n(x)$ is a vector including the
polarization orientation and that we choose them to be real for
convenience.
The electric field operator \cite{note} then reads
\begin{equation}
  \label{eq:efield}
  \hat E(x)=\sum_n \varepsilon_nu_n(x)\left(\hat a_n +\hat
    a_n^\dagger\right),
\end{equation}
where $\hat a_n$ and $\hat a_n^\dagger$ are the usual creation and
annihilation operators of the field excitations and the so-called vacuum
field amplitude is
\begin{equation}
  \varepsilon_n=\sqrt{\frac{\hbar \omega_n}{2\epsilon_0 V}}.
\label{eq:vacfield}
\end{equation}

We assume an amplifier medium consisting of two level atoms.
This model has frequently been used for the quantum treatment of laser
systems \cite{scully}.
The
atoms start in the upper level, interact independently with the field
for a short time before they are repumped incoherently to the upper
level.
The master equation for the density operator of the field is then
obtained by tracing out the atomic variables.
To describe the interaction between the field and a single atom at
position $x$, we
use the interaction Hamiltonian
\begin{equation}
  \label{eq:Hint}
  \hat {\tilde H}=d\sum_n\varepsilon_n u_n(x)
  (\hat a_n^\dagger \hat
  \sigma^- + \hat a_n\hat\sigma^+),
\end{equation}
where $\hat\sigma^\pm$ are the raising and lowering operators for the two
atomic levels
coupled by the dipole moment $d$.
After a short interaction time $\tau$ the change of the reduced field
density operator $\hat{\tilde\rho}(t)$, in the interaction picture, is
given by
\begin{eqnarray}
  \label{eq:deltarho}
  \delta\hat{\tilde\rho}(t)&=&\frac{\tau^2}{2\hbar^2}\sum_{n,m}
  \varepsilon_n\varepsilon_m
  [u_n(x)d][u_m(x)d] e^{i(\omega_n-\omega_m)t}\left\{
    2\hat a_n^\dagger\hat{\tilde\rho}(t)\hat a_m - \hat a_m\hat
  a_n^\dagger\hat{\tilde\rho}(t) - \hat{\tilde\rho}(t)\hat a_m\hat a_n^\dagger
  \right\} \nonumber \\ &&+ O(\tau^3)
\end{eqnarray}
for one atom.
We now assume that the atoms are introduced uniformly distributed over
a volume $V'\neq V$ in the upper level at a
rate $R$.
In principle we could let $R$ depend on the position $x$ and the reservoir
average could include an average over the orientation of the dipole
moment $d$.
Without such refinements, the rate of change of the ensemble averaged density
operator is given by
\begin{eqnarray}
  \label{eq:drho}
  \frac{d}{dt}\hat{\tilde\rho}(t)&=&
  R\frac{1}{V'}\int d^3x\,\delta\hat{\tilde\rho}(t)\nonumber\\
  &=&\frac{1}{2}\sum_{n,m}L_{m,n}e^{i(\omega_n-\omega_m)t}\left\{
    2\hat a_n^\dagger\hat{\tilde\rho}(t)\hat a_m - \hat a_m\hat
  a_n^\dagger\hat{\tilde\rho}(t) - \hat{\tilde\rho}(t)\hat a_m\hat a_n^\dagger
  \right\},
\end{eqnarray}
with the matrix elements
\begin{equation}
  \label{eq:L}
  L_{m,n}=\frac{R\tau^2}{\hbar^2} \varepsilon_n\varepsilon_m
  \frac{1}{V'}\int d^3x\, [u_n(x)d][u_m(x)d].
\end{equation}
Transforming from the interaction picture to the laboratory frame we
get
\begin{equation}
  \label{eq:master}
  \frac{d}{dt}\hat\rho(t)=\frac{1}{2}\sum_{n,m}L_{m,n}\left\{
    2\hat a_n^\dagger\hat\rho(t)\hat a_m - \hat a_m\hat
  a_n^\dagger\hat\rho(t) - \hat\rho(t)\hat a_m\hat a_n^\dagger
  \right\}-i\sum_n\omega_n[\hat a_n^\dagger\hat a_n,\hat\rho(t)].
\end{equation}
This multi-mode master equation with a symmetric matrix $L_{m,n}$
allows us to define the quasi modes and derive their noise properties
in the following. Its important feature is the coupling between modes
with different frequencies due to the reservoir.

We may use the same kind of model also to derive dissipative losses of the
field if we assume a fraction of the reservoir atoms to be initially in the
lower state.
These can be introduced into the system with a distribution differing
from that of the amplifying atoms.
But we will focus now only on the linear amplifier and discuss the case with
damping later.

We are looking for a mode operator $\hat A$ which obeys
\begin{equation}
  \label{eq:qmodeA}
  \frac{d}{dt}\langle \hat A\rangle
  =(\frac{\lambda}{2}-i\Omega)\langle \hat A \rangle
\end{equation}
for an arbitrary field state.
Here $\Omega$ is the real frequency and $\lambda$ is the real
amplification rate.
We write this mode operator in terms of the free field mode
operators as
\begin{equation}
  \label{eq:decomp}
  {\cal E}\hat A=\sum_n\varepsilon_n c_n \hat a_n
\end{equation}
with
the expansion coefficients $c_n$.
This transformation includes the vacuum-field amplitudes
$\varepsilon_n$ and  we define
${\cal E}=\sqrt{\frac{\hbar\Omega}{2\epsilon_0 V}}$,
because then the classical field amplitudes
$\varepsilon_n\langle \hat a_n\rangle$ obey the same transformation.
Inserting Eq.~(\ref{eq:qmodeA}) into (\ref{eq:decomp}) we get an
eigenvalue equation
\begin{equation}
  \label{eq:eigenvalue}
  \sum_n\left( \frac{1}{2}L_{m,n} -i\delta_{n,m}\omega_n \right)
  \frac{\varepsilon_n}{\varepsilon_m} c_n=
  (\frac{\lambda}{2}-i\Omega) c_m
\end{equation}
for the non-Hermitian matrix $\tilde  L_{m,n}=(\frac{1}{2}L_{m,n}
-i\delta_{n,m}\omega_n) \frac{\varepsilon_n}{\varepsilon_m}$.
Here $c_n^{(\nu)}$ is the right eigenvector of $\tilde  L_{m,n}$;
the corresponding left eigenvector is $\varepsilon^2_n c_n^{(\nu)}$
\cite{fn:ev}.
The superscript $\nu$ distinguishes the different eigenvectors.

The only properties of the left and right eigenvectors of
non-Hermitian matrices which we need for our analysis are their
mutual orthogonality and completeness \cite{footnote}:
The eigenvectors fulfill the orthogonality condition
\begin{equation}
  \label{eq:orthogonal}
  \sum_n \varepsilon_n^2
  c^{(\nu)}_nc^{(\mu)}_n=\delta_{\nu,\mu}\sum_n \varepsilon_n^2
  {c^{(\nu)}_n}^2
\end{equation}
and the completeness relation
\begin{equation}
  \label{eq:complete}
  \sum_\nu \left(\frac{
    \varepsilon_n^2 c^{(\nu)}_n c^{(\nu)}_m
    }{
    \sum_{n'} \varepsilon_{n'}^2 {c^{(\nu)}_{n'}}^2}\right)=
  \delta_{n,m}
\end{equation}
with $\sum_{n'} \varepsilon_{n'}^2 {c^{(\nu)}_{n'}}^2\neq0$.
We can now uniquely define the set of quasi-mode operators as
\begin{eqnarray}
  \label{eq:qA}
  \hat A_\nu=\frac{1}{{\cal E}_\nu}\sum_n c^{(\nu)}_n
  \varepsilon_n \hat a_n
\end{eqnarray}
with the vacuum field amplitude
\begin{equation}
  {\cal E}_\nu=\sqrt{\frac{\hbar\Omega_\nu}{2\epsilon_0 V}}.
  \label{eq:qvac}
\end{equation}
The inverse transformation is
\begin{equation}
  \label{eq:a}
  \hat a_n=\varepsilon_n\sum_\nu \frac{c^{(\nu)}_n}{\sum_m \varepsilon_m^2
  {c^{(\nu)}_m}^2}
  {\cal E}_\nu \hat A_\nu.
\end{equation}
Consequently the positive frequency part of the electric field
operator is given by
\begin{eqnarray}
  \label{eq:E-}
  \hat E^{(+)}(x)&=&\sum_n \varepsilon_n u_n(x)\hat a_n\nonumber \\&=&
  \sum_\nu {\cal E}_\nu \left(\sum_n \frac{\varepsilon_n^2
  c^{(\nu)}_n}{\sum_m \varepsilon_m^2
  {c^{(\nu)}_m}^2} u_n(x)\right) \hat A_\nu\nonumber \\&=&
\sum_\nu {\cal E}_\nu U_\nu(x) \hat A_\nu.
\end{eqnarray}
The quasi-mode eigenfunctions
\begin{equation}
  \label{eq:rmode}
  U_\nu(x)=\sum_n \frac{\varepsilon_n^2
  c^{(\nu)}_n}{\sum_m \varepsilon_m^2
  {c^{(\nu)}_m}^2} u_n(x)
\end{equation}
satisfy an orthogonality relation
\begin{equation}
  \label{eq:ortho}
  \frac{1}{V}\int d^3x\, U_\nu(x)\bar U_\mu(x)=\delta_{\nu,\mu}
\end{equation}
with their adjoint quasi-mode functions
\begin{equation}
  \label{eq:lmode}
  \bar U_\nu(x)=\sum_n c^{(\nu)}_n u_n(x).
\end{equation}
The quasi-mode functions have the norm
\begin{equation}
  \label{eq:norm}
  N^2_\nu=\frac{1}{V}\int d^3x\, U_\nu(x)U^*_\nu(x)=
  \frac{\sum_n \varepsilon_n^4 |c^{(\nu)}_n|^2}{\left|\sum_m \varepsilon_m^2
      {c^{(\nu)}_m}^2\right|^2}
\end{equation}
and their adjoints
\begin{equation}
  \label{eq:norma}
  \bar N^2_\nu=\frac{1}{V}\int d^3x\, \bar U_\nu(x)\bar U^*_\nu(x)=
  \sum_n |c^{(\nu)}_n|^2.
\end{equation}

From now on we only consider one quasi mode and drop the index $\nu$.
This is well justified, because, after a long enough time, that quasi
mode which has the largest amplification rate $\lambda$ is dominating
in the sum Eq.~(\ref{eq:E-}). Thus all expectation values can be
calculated with just this largest contribution.

Later we need the properties
\begin{equation}
  \Omega=\frac{\sum_n \varepsilon_n^2\omega_n |c_n|^2}{\sum_n
  \varepsilon_n^2|c_n|^2}=
  \frac{2\epsilon_0 V}{\hbar}
  \frac{\sum_n \varepsilon_n^4 |c_n|^2}{\sum_n
  \varepsilon_n^2|c_n|^2}
  \label{eq:Omega}
\end{equation}
and
\begin{equation}
  \lambda=\frac{\sum_{n,m} L_{n,m}\varepsilon_n\varepsilon_m
  c_n^*c_m}{\sum_n\varepsilon_n^2|c_n|^2}
  \label{eq:lambda}
\end{equation}
which can be obtained from the real and imaginary parts of
Eq.~(\ref{eq:eigenvalue}) after taking the scalar product with the
vector $\varepsilon_m^2c_m^*$.
Note that $\Omega$ is the mean frequency with respect to the
probabilities
\begin{equation}
  p_n=\frac{\varepsilon_n^2|c_n|^2}{\sum_{n'}\varepsilon_{n'}^2|c_{n'}|^2}.
  \label{eq:prob}
\end{equation}

We calculate now the noise of the slowly-varying quadrature operator
\begin{equation}
  \label{eq:X}
  \hat X(x)=
  {\cal E}
  (U(x)\hat A e^{i\Omega t}+U^*(x)\hat A^\dagger e^{-i\Omega t})
\end{equation}
in a frame rotating with $\Omega$.
After some straightforward calculation we get the time evolution of the noise
\begin{equation}
  \label{eq:noise}
  \frac{d}{dt}(\Delta X(x))^2=\lambda(\Delta X(x))^2+\lambda
  |U(x)|^2 \sum
  \varepsilon_n^2|c_n|^2.
\end{equation}
We have to compare this with the noise of the usual
single-mode-amplifier master equation
\begin{equation}
  \label{eq:usualmaster}
  \frac{d}{dt}\hat\rho(t)=\frac{\lambda}{2}
  \left\{2\hat a^\dagger\hat\rho(t)\hat a
  -\hat a\hat a^\dagger\hat\rho(t)-\hat\rho(t)\hat a\hat a^\dagger\right\}
  -i\Omega[\hat a^\dagger\hat a,\hat\rho(t)]
\end{equation}
for the mode frequency $\Omega$, normalized mode function $u(x)$, and
the vacuum field amplitude
${\cal E}=\sqrt{\frac{\hbar\Omega}{2\epsilon_0 V}}$.
In this case the noise is given by the equation
\begin{equation}
  \frac{d}{dt}(\Delta X(x))^2=\lambda(\Delta
  X(x))^2+\lambda |u(x)|^2{\cal E}^2.
  \label{eq:usualnoise}
\end{equation}
Hence, after averaging over position in Eqs.~(\ref{eq:noise}) and
(\ref{eq:usualnoise}) and using Eq.~(\ref{eq:norm}), we can read off
the excess-noise factor
\begin{eqnarray}
  \label{eq:Kqm}
  K&=&\frac{N^2}{{\cal E}^2}\sum_n \varepsilon_n^2|c_n|^2=\frac{\sum
      \varepsilon_n^2|c_n|^2 \sum_n \varepsilon_n^4
      |c_n|^2}{{\cal E}^2\left|\sum_m \varepsilon_m^2
      {c_m}^2\right|^2}\nonumber \\&=&
  \left|\frac{\sum\varepsilon_n^2|c_n|^2}{\sum_m \varepsilon_m^2
      {c_m}^2}\right|^2.
\end{eqnarray}
From the triangular inequality $\sum\varepsilon_n^2|c_n|^2 \ge
|\sum_m\varepsilon_m^2{c_m}^2|$ follows $K\ge 1$. The peculiar mode
coupling of the non-Hermitian eigenvalue equation
(\ref{eq:eigenvalue}) causes this enhancement of the reservoir noise
entering the quasi mode.

We can now compare the result, Eq.~(\ref{eq:Kqm}), with the expression
\begin{eqnarray}
  \label{eq:KK}
  \tilde K&=&\frac{\int d^3x\,U(x)U^*(x) \int d^3x\,\bar U(x) \bar
    U^*(x)}{\left|\int
      d^3x\,U(x)\bar U(x)\right|^2}\nonumber\\&=&
  N^2\bar N^2=
  \frac{\sum_n
    |c_n|^2 \sum_n \varepsilon_n^4|c_n|^2}{\left|\sum_m \varepsilon_m^2
    {c_m}^2\right|^2}
\end{eqnarray}
which has been derived from a
semi-classical theory \cite{semi}. We see that they are almost identical
when we use
Eqs.~(\ref{eq:ortho}), (\ref{eq:norm}) and (\ref{eq:norma}).
The ratio of the factors is given by
\begin{eqnarray}
  \label{eq:K-K}
  \frac{\tilde K}{K}&=&\frac{ {\cal E}^2\sum_n |c_n|^2}{\sum_n
    \varepsilon_n^2|c_n|^2 }=
    \frac{\sum_n|c_n|^2\sum_n|c_n|^2\varepsilon_n^4}{
    (\sum_n|c_n|^2\varepsilon_n^2)^2} \nonumber\\&=&
    \Omega\overline{\left(1/\omega\right)}\approx 1
    +{\cal O}(\Delta \omega/\Omega)^2
\end{eqnarray}
with the mean inverse frequency $\overline{\omega^{-1}}$ with respect
to the probabilities $p_n$ introduced above in Eq.~(\ref{eq:prob}).
Here we have used the definitions of $\varepsilon_n$, ${\cal E}$ and
$\Omega$ in Eqs.~(\ref{eq:vacfield}), (\ref{eq:qvac}) and
(\ref{eq:Omega}). The quantity $\Delta\omega$ is a measure of the
bandwidth of the quasi mode in terms of the modes of the universe.
From Eqs.~(\ref{eq:Kqm}) and (\ref{eq:KK}) we see, by using Schwarz'
inequality, that $K\le\tilde K$. The correction to the semi-classical
result is small for the optical frequency domain where the bandwidth
($<10^{10}$ Hz) is negligible with respect to the mean frequency
($>10^{14}$ Hz). But in the micro-wave regime (GHz) the correction may
be essential.

We note that if the quasi-mode frequency $\Omega$ was defined with respect
to the weights $|c_n|^2$ instead of Eq.~(\ref{eq:prob}), the ratio
Eq.~(\ref{eq:K-K}) would be unity.
Indeed, $|c_n|^2$ is the weight of the classical field amplitudes
$\varepsilon_n \langle \hat a_n\rangle$ in Eq.~(\ref{eq:decomp}) and it is
the weight of the universe modes $u_n(x)$ in the adjoint quasi-mode functions
$\bar U_\nu(x)$ in Eq.~(\ref{eq:lmode}).
On the other hand, the weight of the universe modes in the quasi-mode
functions in Eq.~(\ref{eq:rmode}) is $\varepsilon_n^4|c_n|^2$.
But, Eq.~(\ref{eq:Omega}) shows that the geometric mean of the two
classical possibilities gives the proper distribution $p_n$ for the
frequencies with the eigenfrequency $\Omega$ as mean value.

As a next step we could derive damping for the quasi modes by
considering the limit of a continuum of modes of the universe. In the
case of classical field modes, this was done by Lang, Scully and Lamb
\cite{lang} in one dimension for a cavity with one perfect and one
semi-transparent mirror. It is interesting to note that in this
special case, the authors could explicitly prove that only one quasi
mode exhibits amplification whereas all other quasi modes experience
attenuation. However, in the context of discussing the properties of
excess noise in this paper, we are satisfied by adding damping due to
a separate reservoir. In an analogous way as before we can derive the
multi-mode master equation
\begin{eqnarray}
  \label{eq:masterda}
  \frac{d}{dt}\hat\rho(t)&=&\frac{1}{2}\sum_{n,m}L_{m,n}\left\{
    2\hat a_n^\dagger\hat\rho(t)\hat a_m - \hat a_m\hat
    a_n^\dagger\hat\rho(t) - \hat\rho(t)\hat a_m\hat a_n^\dagger
  \right\}\nonumber\\
  &&+\frac{1}{2}\sum_{n,m}\Gamma_{m,n}\left\{
    2\hat a_n\hat\rho(t)\hat a_m^\dagger - \hat a_m^\dagger\hat
    a_n\hat\rho(t) - \hat\rho(t)\hat a_m^\dagger\hat a_n
  \right\}
  -i\sum_n\omega_n[\hat a_n^\dagger\hat a_n,\hat\rho(t)].
\end{eqnarray}
The symmetric matrix $\Gamma_{m,n}$ is defined similar to the matrix
$L_{m,n}$ in Eq.~(\ref{eq:L}), with a possibly different volume $V'$
and a possibly different atom injection rate $R$ for the lower state.

We are interested in the case when the strength of the amplification
is adjusted such that the quasi-mode amplitude obeys the equation
\begin{equation}
  \label{eq:defAda}
  \frac{d}{dt}\langle \hat A \rangle=-i\Omega\langle \hat A \rangle
\end{equation}
yielding a stationary modulus.
At threshold, the laser quasi-mode is oscillating without change of
amplitude.
Thus, the effects of amplification and attenuation must compensate
exactly on the average, but only  for this quasi-mode. This does not
mean that the two contributions cancel at the level of the master
equation, but only that we can find an eigenmode with a purely
imaginary eigenvalue.
In fact, assuming exact cancellation in the master equation leads us to
a trivial case, see below.
In Ref.\ \cite{gold} it is assumed that the same applies to the
saturated gain at steady state operation above threshold; here we
consider only the linear behavior close to threshold.
We return to the discussion of the nonlinear problem in a forthcoming
publication.

These assumptions lead to the eigenvalue problem
\begin{equation}
  \label{eq:eigenvalueda}
  \sum_n\left(\frac{1}{2} L_{m,n}
    -\frac{1}{2}\Gamma_{m,n}-i\delta_{n,m}\omega_n \right)
  \frac{\varepsilon_n}{\varepsilon_m} c_n= -i\Omega c_m.
\end{equation}
We thus require the amplification rate
$\lambda$ as given in Eq.~(\ref{eq:lambda})
and
the damping rate
\begin{equation}
  \gamma=\frac{\sum_{n,m}
    \Gamma_{n,m}\varepsilon_n\varepsilon_m
    c_n^*c_m}{\sum_n\varepsilon_n^2|c_n|^2}
  \label{eq:gamma}
\end{equation}
to be equal; $\gamma=\lambda$.
Proceeding as before and averaging over position, we obtain the equation
\begin{equation}
  \label{eq:noiseda}
  \frac{d}{dt}(\Delta X)^2=(\gamma+\lambda) {\cal E}^2K=2\lambda
  {\cal E}^2K
\end{equation}
for the noise with the same excess-noise factor $K$ as in
Eq.~(\ref{eq:Kqm}).
This describes a diffusion process with the diffusion constant $2D_X=2
\lambda {\cal E}^2K$.
The linewidth of the laser at threshold \cite{schawlow,scully,lax} is
then given by the phase diffusion constant $2D_\phi=2D_X/I=K\lambda^2/4P$
with field intensity $I=4{\cal E}^2N^2\langle \hat
A^\dagger\hat 
A\rangle$ and output power $P=\gamma I/(4{\cal
  E}^2)=\gamma N^2 
\langle \hat A^\dagger\hat A\rangle$ measured in photon energies.
The measurement of the laser intensity is not straightforward in the
case of a quasi-mode.
However, if the detector is mode matched to the outgoing quasi-mode
profile, the maximum intensity $I$ is obtained.
All other detector arrangements will miss some intensity and give a
lower value.
In the present case, we find the noise to be purely due to phase
diffusion.
This derives from our assumption that we consider the steady state
near threshold, not a linearized noise theory around the saturated
gain. In the latter case, we additionally would expect a line width
contribution from the amplitude fluctuations.

We consider now the steady state case when the mode coupling is mainly
caused by the damping, which holds if $L_{m,n}\approx\lambda
\delta_{m,n}$.
Then we get from Eq.~(\ref{eq:eigenvalueda}) the eigenvalue equation
\begin{equation}
  \label{eq:eigenvaluedae}
  \sum_n\left( -\frac{1}{2}\Gamma_{m,n}-i\delta_{n,m}\omega_n \right)
  \frac{\varepsilon_n}{\varepsilon_m} c_n= (-\frac{\gamma}{2}-i\Omega) c_m
\end{equation}
which is the equation describing a system without an amplifying
medium.
Since with this assumption the eigenvectors are the same for
Eqs.~(\ref{eq:eigenvalueda}) and (\ref{eq:eigenvaluedae}), also the
$K$-factor is the same in both cases.
This result justifies the commonly used procedure to calculate
the $K$-factor of a laser system from the cavity-decay properties
alone.
Our assumption probably gives an over-simplified picture of a real
laser.
However, in most systems the gain medium is distributed uniformly over
an essential part of the mode volume.  
From the definition, Eq.~(\ref{eq:L}), then follows that
$L_{m,n}\approx\lambda \delta_{m,n}$.
Besides this, the gain is often due to a resonant interaction whereas the
damping is highly broad band.
In that situation it seems reasonable to assume the gain to be
diagonal in the ``universe'' modes and the mode-mode coupling to derive
mainly from the loss mechanisms.
These contain diffractive losses at the cavity edges and transmission
losses, and hence their distribution is bound to differ greatly from
the distribution of the amplification.
In comparison to the off-diagonal elements of $\Gamma_{m,n}$, the
off-diagonal elements of $L_{m,n}$ can be neglected.

From Eq.~(\ref{eq:eigenvalueda}) we can also see in which case the
minimum excess noise $K=1$ can be achieved.
For $L_{m,n}=\Gamma_{m,n}$ we have $\hat a_n=\hat A_n$.
Losses and
amplification are compensated in each volume element.
This means that, when damping and amplification are acting equally in
the same volume $V'$, the true modes and the quasi modes are
identical.
This can happen when losses are mainly provided by the same entities
which give the gain like in dye lasers.
In most lasers, however, it is  mainly the smallness of the
off-diagonal elements of the matrices $L_{m,n}$ and $\Gamma_{m,n}$
which leads to a $K$-factor close to one.

Summarizing our main results, we have derived the excess-noise factor
$K$ within the framework of the conventional quantum theory of laser-like
systems.
Our calculation is conceptually simple and
our result also shows a small correction to the well established
semi-classical one.

Acknowledgement. We thank M.\ T.\ Fontenelle and U.\ Leonhardt for
helpful discussion.
We also thank A.\ E.\ Siegman and J.\ P.\ Woerdman for providing
manuscripts prior to publication.
One of us (P.J.B.) thanks the Alexander von Humboldt
Foundation for
supporting his work at the Royal Institute of Technology.


\begin{thebibliography}{99}
\bibitem{note} In principle we should define the magnetic field
  operator as well. This would require an additional set of magnetic
  mode functions related to the electric mode functions by a curl
  operation. But here we only need to consider the electric field.
\bibitem{fn:ev} For the case of a symmetric matrix $L_{m,n}$, the
  relation between left and right eigenvectors is easy to see.
  However, a similar relation holds also in the case of complex mode
  functions $u_n(x)$, when the matrix $L_{m,n}$ is Hermitian.
  The left eigenvectors are then given by
  $\sum_m M_{n,m} \varepsilon^2_m c^{(\nu)}_m$ with the unitary matrix
  $M_{n,m}$.
\end{thebibliography}
\end{document}